\documentclass[sigconf]{acmart}
\AtBeginDocument{%
  \providecommand\BibTeX{{%
    \normalfont B\kern-0.5em{\scshape i\kern-0.25em b}\kern-0.8em\TeX}}}

\copyrightyear{2024} 
\acmYear{2024} 
\setcopyright{acmlicensed}
\acmConference[KDD '24]{Proceedings of the 30th
ACM SIGKDD Conference on Knowledge Discovery and Data Mining}{August
25--29, 2024}{Barcelona, Spain}
\acmBooktitle{Proceedings of the 30th ACM SIGKDD Conference on Knowledge
Discovery and Data Mining (KDD '24), August 25--29, 2024, Barcelona, Spain}
\acmDOI{10.1145/3637528.3671787}
\acmISBN{979-8-4007-0490-1/24/08}
\usepackage{latexsym}
\usepackage{multirow}
\usepackage{booktabs}
\usepackage{subfigure}
\usepackage{amsfonts}
\usepackage{amsthm}

\usepackage{tablefootnote}
\usepackage{float}

\definecolor{commentcolor}{RGB}{110,154,155}   % define comment color
\begin{document}

\title{Improving the Consistency in Cross-Lingual Cross-Modal Retrieval with 1-to-K Contrastive Learning}

\author{Zhijie Nie}
\affiliation{%
  \institution{CCSE, Beihang University}
  \state{Beijing}
  \country{China}}
\email{niezj@act.buaa.edu.cn}

\author{Richong Zhang}
\authornote{Corresbonding author: zhangrc@act.buaa.edu.cn.}
\affiliation{%
  \institution{CCSE, Beihang University}
  \state{Beijing}
  \country{China}}
\email{zhangrc@act.buaa.edu.cn}

\author{Zhangchi Feng}
\affiliation{%
  \institution{CCSE, Beihang University}
  \state{Beijing}
  \country{China}}
\email{zcmuller@buaa.edu.cn}

\author{Hailang Huang}
\affiliation{%
  \institution{CCSE, Beihang University}
  \state{Beijing}
  \country{China}}
\email{huanghl@act.buaa.edu.cn}

\author{Xudong Liu}
\affiliation{%
  \institution{CCSE, Beihang University}
  \state{Beijing}
  \country{China}}
\email{liuxd@act.buaa.edu.cn}
\renewcommand{\shortauthors}{Zhijie Nie, Richong Zhang, Zhangchi Feng, Hailang Huang, \& Xudong Liu}
% \renewcommand{\shortauthors}{Anonymous}
\begin{abstract}
Cross-lingual Cross-modal Retrieval (CCR) is an essential task in web search, which aims to break the barriers between modality and language simultaneously and achieves image-text retrieval in the multi-lingual scenario with a single model. In recent years, excellent progress has been made based on cross-lingual cross-modal pre-training; particularly, the methods based on contrastive learning on large-scale data have significantly improved retrieval tasks. However, these methods directly follow the existing pre-training methods in the cross-lingual or cross-modal domain, leading to two problems of inconsistency in CCR: The methods with cross-lingual style suffer from the intra-modal error propagation, resulting in inconsistent recall performance across languages in the whole dataset. The methods with cross-modal style suffer from the inter-modal optimization direction bias, resulting in inconsistent rank across languages within each instance, which cannot be reflected by Recall@K. To solve these problems, we propose a simple but effective 1-to-K contrastive learning method, which treats each language equally and eliminates error propagation and optimization bias. In addition, we propose a new evaluation metric, Mean Rank Variance (MRV), to reflect the rank inconsistency across languages within each instance. Extensive experiments on four CCR datasets show that our method improves both recall rates and MRV with smaller-scale pre-trained data, achieving the new state-of-art\footnote{Our codes can be accessed at \textcolor{blue}{\url{https://github.com/BUAADreamer/CCRK}}}.
\end{abstract}





%%
%% This command processes the author affiliation and title
\maketitle

\section{Introduction}

\begin{figure}[tp]
    \centering
    \includegraphics[width=\linewidth]{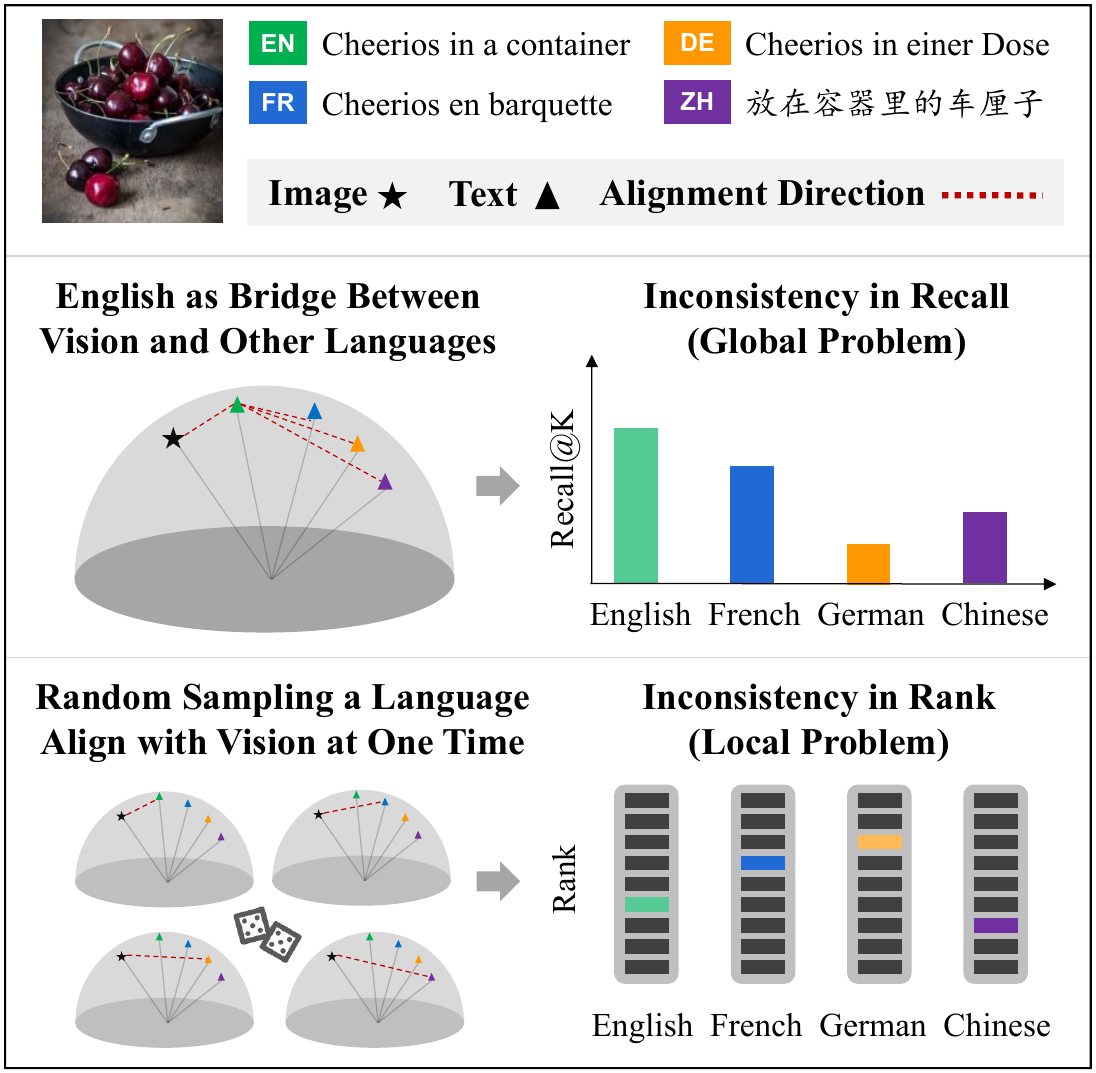}
    \caption{Two inconsistency problems exist in the current cross-lingual cross-modal pre-training methods, leading to inconsistent recall and ranking in cross-lingual cross-modal retrieval separately.}
    \label{fig:problem_impact}
\end{figure}

Recently, significant progress has been made in the cross-modality \cite{radford2021learning,li2021align,su2022towards}, and the cross-lingual \cite{devlin2018bert,conneau2019cross,chi2021infoxlm} domains, leading to increased interest in the more general cross-lingual cross-modal scenarios.  In the cross-lingual cross-modal domain, Cross-lingual Cross-modal Pre-training (CCP) \cite{ni2021m3p,zhou2021uc2,shan2022ernie,zeng2022cross} is first explored, followed by Cross-lingual Cross-modal Retrieval (CCR) \cite{portaz2019image,jain2021mural,fei2021cross,carlsson2022cross,wang2022cross} as the first downstream task independently studied. CCR aims to achieve image-text retrieval in multi-lingual scenarios with a single model, preventing the high latency associated with text translation from other languages to English in real-time web searches.

In general, modern dense retrieval matches the results for a query by a particular distance metric (e.g., Euclidean distance or cosine similarity), which implies that the dense retrieval methods should push queries and those semantically similar candidate items closer than other random pairs in the high-dimensional space. Thus, the core of the retrieval task lies in aligning the semantic spaces of queries and candidate sets, regardless of whether they are in different languages or different modalities. Recent studies show that contrastive learning based on pairwise data is effective in cross-lingual and cross-modal retrieval tasks. For example, CLIP \cite{radford2021learning}, which is only pre-trained by aligning different modalities using contrastive learning, has achieved remarkable performances in zero-shot cross-modal retrieval; on the other hand, aligning the representations from different modalities (or different languages) before fusing them can reduce the difficulty of fusion and significantly improve the performance of downstream cross-modal tasks including retrieval, question answering and reasoning \cite{li2021align}. As a result, the existing works in CCP directly pieced the alignment ideas in cross-modal or cross-lingual domains, feeding pairwise data into the encoder at a time, such as an image-text pair and a bi-lingual text pair. Specifically, the existing methods use the following two ideas to align different modalities: (1) considering English as the anchor for bridging vision with other languages, which means that the images are aligned to the English texts only, while the texts in other languages are aligned to the English texts only \cite{jain2021mural,zeng2022multi} or (2) considering the images being aligned with the texts in a random language at a time during pre-training \cite{zeng2022cross}.

However, the desirable alignment process is more complex in cross-lingual cross-modal scenarios. Intuitively, the semantics of the texts in multiple languages need to be aligned jointly with those from vision, which cannot be achieved with pairwise data. With the theoretical derivations and empirical studies (Section \ref{sec:problem}), we find that applying either of the two above ideas to CCP will result in two problems of inconsistency (Figure \ref{fig:problem_impact}). Specifically, regarding English as the bridge in inter-modal may cause error propagation, resulting in an inconsistent performance on Recall@K of different languages in CCR; aligning the image with only the text in a random language at a time may lead to the optimization direction bias, resulting the inconsistent ranks of different languages within an instance. Highlighting that the latter problem is more insidious since it cannot be directly reflected by Recall@K, which is almost the only reported evaluation metric of CCR \cite{ni2021m3p,zhou2021uc2,bugliarello2022iglue,zeng2022cross}.

To solve the above problems, in this paper, we propose a simple but effective contrastive paradigm for CCP, 1-to-K contrastive learning. Specifically, when pre-training the images and texts in a mini-batch ratio of not 1 to 1 but 1 to K (K $\geq$ 2), each image is aligned simultaneously with K texts in different languages. Under this paradigm, all languages are aligned with vision at once, and no language is used as the bridge between vision and other languages, eliminating intra-modal error propagation and inter-modal optimization direction bias in principle. In addition, two commonly used pre-training tasks for capturing fine-grained correlation between modalities, Multi-lingual Image-Text Matching (MITM) \cite{zhou2021uc2,zeng2022cross} and Cross-modal Masked Language Modeling (CMLM) \cite{ni2021m3p,zeng2022cross}, can be easier superimposed on the novel contrastive paradigm with the help of hard negative sampling.  Based on the three pre-training tasks, we propose a pre-trained model, CCR$^k$. For the evaluation of CCR, as a complement to Recall@K, we propose a new evaluation metric, Mean Rank Variance (MRV), to reflect the rank inconsistency of the different languages in an instance. Extensive experiments on four public CCR datasets demonstrate that our method has effectively solved the above two problems and achieved new state-of-the-art.

The contributions of this paper can be summarized as follows:
\begin{itemize}
    \item We analyze two problems of inconsistency existing in the current CCP methods and point out their impact on the performance of CCR for the first time.
    \item We propose a simple but effective 1-to-K contrastive paradigm as an alternative to the traditional 1-to-1 contrastive paradigm in CCR to solve these problems.
    \item We propose Mean Rank Variance (MRV) to better reflect retrieval performance across languages and modalities, which is used to replenish Recall@K and evaluate the rank consistency across languages in each dataset sample.
    \item  We propose CCR$^k$, a CCP model with the novel 1-to-K contrastive paradigm. We pre-train four variants of CCR with the different language numbers and data scales. The largest variant CCR$^{10}$-E, which is still pre-trained with fewer language numbers and data scale than all baselines, achieves new SOTA on four CCR datasets.
\end{itemize}

\section{Background}
This section overviews recent advances in cross-lingual cross-modal pre-training and cross-lingual cross-modal retrieval. Due to space limitations, we will only focus on works related to image-text retrieval in the cross-lingual scenarios.

\subsection{Cross-Lingual Cross-Modal Pre-Training}\label{sec:ccp}
Cross-lingual Cross-modal Pre-training (CCP) \cite{ni2021m3p,zhou2021uc2,shan2022ernie,zeng2022cross} is generalized from cross-modal pre-training \cite{li2021align,bao2022vlmo,su2022towards} and cross-lingual pre-training \cite{devlin2018bert,conneau2019cross,chi2021infoxlm}, which aims to develop a representation learning model that captures the relationship in different modalities and different languages simultaneously. Current methods can be broadly divided into three categories based on their model architectures.

\paragraph{\bf Cross-Lingual Style} The first class of methods follows the model architecture in the cross-lingual domain, where a pre-trained cross-modal model (e.g. CLIP \cite{radford2021learning}) is required. Then, the pre-trained model is tuned to a cross-lingual version by aligning the representations of English texts and non-English texts while freezing both the visual and English textual backbone. The representatives of these methods are multi-lingual CLIPs \cite{carlsson2022cross, tyshchuk2023isotropy}. The idea behind these methods is {\bf using English as a bridge between vision and other languages}.

\paragraph{\bf Cross-Modal Style} The second class of methods follows the model architecture in the cross-modal domain, where multi-lingual image-text pairs are required. Due to the difficulty of collecting multi-lingual image-text pairs in practice, translation models are usually used to translate the English text in the existing image-text pairs to other languages \cite{zhou2021uc2,qiu2022,zeng2022cross,jain2021mural}. Then, at most, one non-English text is adapted to form an image-text pair with the image at a time, keeping consistent with the input form of the cross-modal model \cite{li2021align,radford2021learning}. The representatives of these methods are UC$^2$ \cite{zhou2021uc2}, and TD-MML \cite{qiu2022}. The idea behind these methods is {\bf aligning the image with the text in a language at a time to improve the performance across languages}.

\paragraph{\bf Cross-Modal Cross-Lingual Style} The third class of methods references the architectures in both cross-lingual and cross-modal domains. The same multi-lingual encoders are responsible for encoding the texts in both image-text pairs and parallel corpora for a unified framework. The representatives of these methods are xUNITER \cite{liu2021visually}, M$^3$P \cite{ni2021m3p}, and CCLM \cite{zeng2022cross}. The idea behind these methods is {\bf using a unified framework to combine the ideas from the first and second class of methods}.

\subsection{Cross-Lingual Cross-Modal Retrieval}
Cross-lingual Cross-modal Retrieval (CCR) \cite{portaz2019image,jain2021mural,fei2021cross,carlsson2022cross,wang2022cross}  is one of the downstream tasks that have been focused on in cross-lingual cross-modal scenarios. MURAL \cite{jain2021mural} demonstrates that high performance in CCR can be achieved through pre-training with contrastive learning over large-scale datasets. \citet{fei2021cross} pre-train only a fusion encoder for CCR using pre-extracted image region features. More recently, IGLUE \cite{bugliarello2022iglue}, a cross-lingual cross-modal benchmark, was proposed with two new retrieval datasets, xFlickr\&CO and WIT. In addition, IGLUE explores several cross-modal pre-training models (such as ViLBERT \cite{lu2019vilbert} and xUNITER \cite{liu2021visually}), and evaluates them on two new datasets by directly translating the texts in other languages to English, demonstrating that these models serve as strong baselines. \citet{carlsson2022cross} apply cross-lingual teacher learning to transfer CLIP to other languages. \citet{wang2022cross} proposed a noise robustness CCR method to improve the performance when training on the noisy translated data.

To the best of our knowledge, our work in this paper is the first exploration of the consistency in cross-lingual cross-modal retrieval. In addition, our newly proposed 1-to-K contrastive learning pre-training task and the evaluation metric MRV have not previously appeared in CCR and related fields.

\section{Problem of Inconsistency in CCR}\label{sec:problem}
In this section, we first explore two alignment problems in the existing CCP methods under the perspective of contrastive learning, then point out their impacts on the performance of CCR.

\subsection{Preliminary}
In the loss functions for alignment, there may be only the anchor with its positive samples (e.g., Mean Squared Error (MSE)) and the optional negative samples (e.g., InfoNCE Loss \cite{oord2018representation}, which is commonly used in contrastive learning). When these loss functions are used, the anchor is optimized by the alignment direction, which points from the anchor to the positive sample. Intuitively, the alignment direction brings the anchor and positives together in the semantic space.

In advance, we give the required notation for the follow-up content in this section. For simplicity, we only consider the case where one image needs to be aligned with two texts from two different languages, and the subsequent conclusions can be easily generalized to more languages. Let $\hat{i}$, $\hat{t}_{m}$ and $\hat{t}_{n}$ denote the normalized representations of the image, the text in language $m$, and the text in language $n$, respectively. We define $\alpha=\angle(\hat{i},\hat{t}_{m})$, $\beta=\angle(\hat{i},\hat{t}_{n})$ and $\gamma=\angle(\hat{t}_{m},\hat{t}_{n})$, where $\angle(.,.)$ represents the angle of two same dimensional representations.

\subsection{Inconsistency in Recall@K}\label{sec:error_pro}

\begin{figure}[htb]
\centering  %图片全局居中
\subfigure[]{
\label{fig:demo1}
\includegraphics[width=0.43\linewidth]{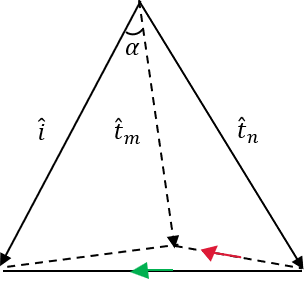}}
\subfigure[]{
\label{fig:emprical_loss}
\includegraphics[width=0.54\linewidth]{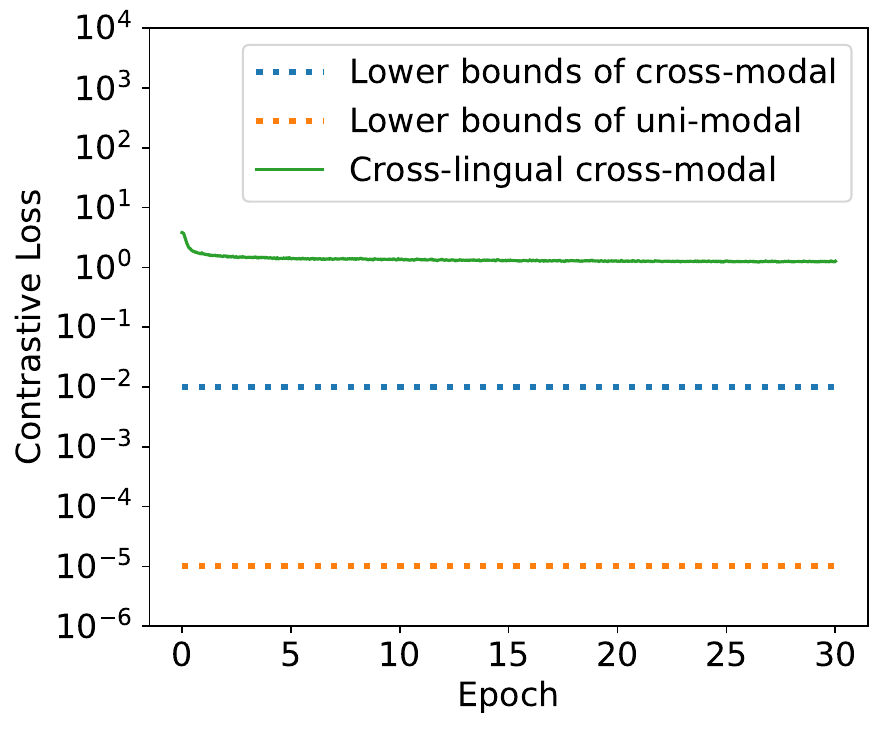}}
\caption{Theoretical analysis and empirical observation for inconsistency in Recall@K. (a) An illustration of Lemma \ref{lemma1}, where the green arrow represents the correct alignment direction, while the red arrow represents the practical alignment direction. (b) A comparison of infoNCE loss value in different scenarios. We pre-trained and recorded loss changes using SimCSE \cite{gao2021simcse} in the uni-modal setting, ALBEF \cite{li2021align} in the cross-model setting and CCLM \cite{zeng2022cross} in CCP, respectively, while keeping other settings as identical as possible.} 
\label{fig:consistency_in_recall}
\end{figure}

{\bf Theoretical Analysis.} The methods following the cross-lingual architecture implicitly rely on English as a bridge in inter-modal alignment between the other language and vision. In this setting, we consider the situation in which the other language text representation is the anchor, where it is aligned to its positive sample, the English text representation. However, in theory, it should be aligned to the image representation. Without loss of generality, if we regard language $m$ as English and language $n$ as another language, then the practical alignment direction is $\hat{t}_{m}-\hat{t}_{n}$, while correct alignment direction is $\hat{i}-\hat{t}_{n}$ (Figure \ref{fig:demo1}). Then we have the following results:
\begin{lemma}\label{lemma1}
Suppose that $\theta$ is the angle between the practical and correct alignment direction of $\hat{t}_{n}$. If and only if English texts can be aligned well with images, i.e. $\alpha$ tends to 0, then $\theta$ will converge to 0.
\end{lemma}

{\bf Empirical Observation.} We find the inter-modal alignment process so tough that English texts cannot be aligned well with images. Specifically, the loss value can drop by 5 to 6 orders of magnitude in the text-modal (uni-modal) scenario \cite{gao2021simcse}, while it is only 2 orders of magnitude in cross-modal contrastive learning \cite{li2021align} (Figure \ref{fig:emprical_loss}). It means that the alignment between English texts and images is not ideal, and if English texts are used to connect images and texts in other languages, there will be a risk of error propagation on intra-modal alignment, resulting in a worse alignment between non-English texts and images.

{\bf Impact of inconsistency.} As this problem persists during pre-training, the impact of this problem is global and can be revealed by the uneven performance under the different language settings. As it is shown by the results of M$^3$P and UC$^2$ in Table \ref{tab:performance}, the performance gap among different language scenarios is clear even though the instance number per language has been kept nearly consistent during pre-training \cite{zhou2021uc2}.

\subsection{Inconsistency in Rank}\label{sec:direction_bias}

{\bf Theoretical Analysis.} The methods that follow the cross-modal architecture consider each language separately aligned to the vision, thus avoiding error propagation in intra-modal. However, they suffer from another local problem of inconsistency.

In this setting, we consider the situation that the image is the anchor, where its optimal alignment coordinates should satisfy: (1) $\min (\angle(\hat{i},\hat{t}_m)+\angle(\hat{i},\hat{t}_n))$ and (2) $\angle(\hat{i}, \hat{t}_m) = \angle(\hat{i}, \hat{t}_n)$. Combining the two conditions above, $\hat{i}$ should be drawn to the midpoint of the minor arc corresponding to $\hat{t}_m$ and $\hat{t}_n$, i.e., the correct alignment direction is $\frac{(\hat{t}_{m}+\hat{t}_{n})}{\|\hat{t}_{m}+\hat{t}_{n}\|}-\hat{i}$.

\begin{figure}[thb]
\centering  %图片全局居中
\subfigure[]{
\label{fig:demo2}
\includegraphics[width=0.43\linewidth]{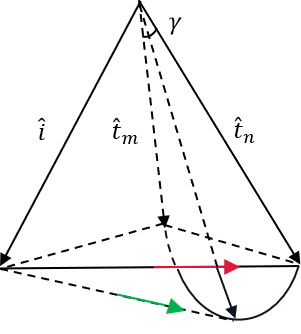}}
\subfigure[]{
\label{fig:emprical_sne}
\includegraphics[width=0.48\linewidth]{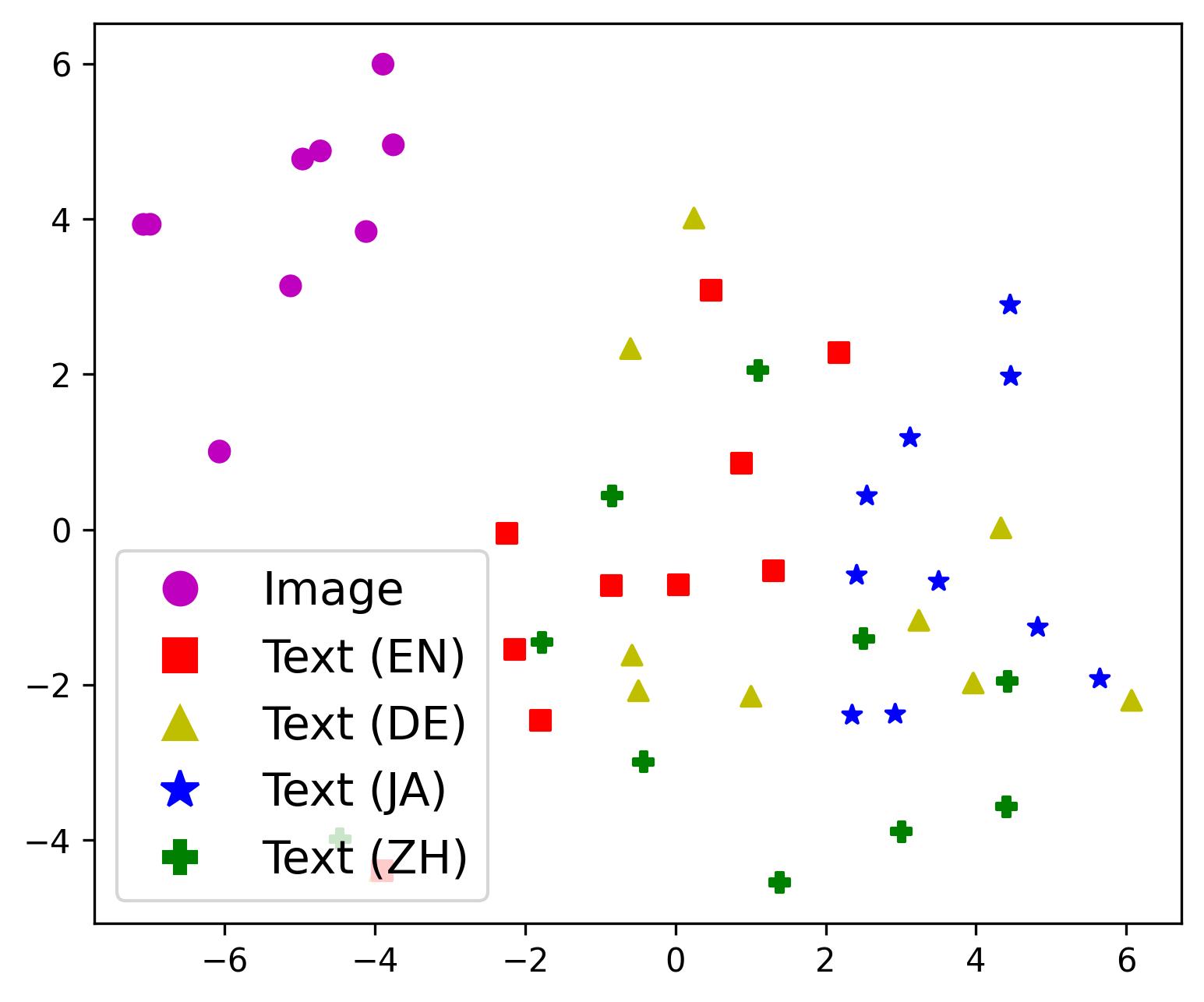}}
\caption{Theoretical analysis and empirical observation for inconsistency in Rank. (a) An illustration of Lemma \ref{lemma2}, where the green arrow represents the correct alignment direction, while the red arrow represents the practical alignment direction. (b) A Visualization of T-SNE with 10 instances randomly sampled in xFlickr\&CO. The representations are obtained by Swin Transformer \cite{liu2021swin} and the first half (first six layers) of XLM-R \cite{conneau2020unsupervised} following the setting in CCLM \cite{zeng2022cross}.} 
\label{fig:consistency_in_rank}
\end{figure}

However, the image is aligned with only one of the text representations at a time under the cross-modal setting. Without loss of generality, if we regard $\hat{t}_{m}$ as the alignment target, the practical alignment direction of $\hat{i}$ can be considered as $\hat{t}_{n}-\hat{i}$ (Figure \ref{fig:demo2}). Then we have the following results:

\begin{lemma}\label{lemma2}
Suppose that $\omega$ is the angle between the actual alignment direction and the correct optimization direction of $\hat{i}$. If and only if the English text can be aligned well with the text in the other language, i.e. $\gamma$ tends to 0, then $\omega$ will converge to 0.
\end{lemma}

{\bf Empirical Observation.} We find that the representations obtained by the popular multi-lingual text encoders are not aligned according to semantics after degenerating the representations by t-SNE \cite{van2008visualizing}. Instead, they remain irregularly distributed (Figure \ref{fig:emprical_sne}). As a result, the alignment direction of the image may not favor all languages when the model only sees the texts in one language at one time, which might result in inconsistent performance among the semantically similar texts in different languages. 

{\bf Impact of inconsistency.} As this problem appears dynamically in different instances for different languages during pre-training, the impact of this problem is local. The very different retrieval results will be obtained (1) when the texts in different languages are retrieved using the same image or (2) when the same image is retrieved using the texts in different languages but with the same semantics. Unfortunately, Recall@K can only reflect the overall performance of the model on each language in the whole dataset but can not reflect the inconsistent performance across languages of an instance.

\begin{figure*}[tp]
    \centering
    \includegraphics[width=\textwidth]{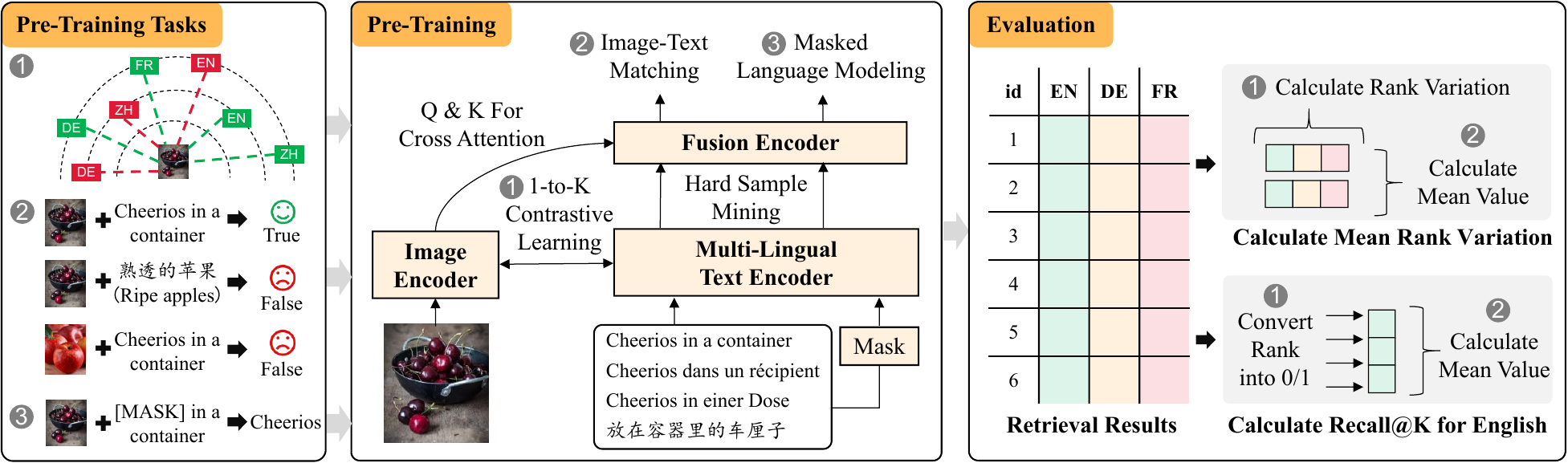}
    \caption{The overview of our pre-training tasks, model architecture, and evaluation metrics.}
    \label{fig:model}
\end{figure*}

\section{Method}\label{sec:method}
The section is organized as follows: some necessary notations are first introduced in Section \ref{sec:model}; a novel 1-to-K contrastive method is then proposed to solve the inconsistency problems in Section \ref{sec:cl}; a pre-training model, CCR$^k$, is further presented to combine 1-to-K contrastive learning with other common pre-training tasks in a unified framework in Section \ref{sec:pretraining}; Finally, a new evaluation metric called Mean Rank Variance (MRV) is proposed in Section \ref{sec:mrv}, which evaluates the rank consistency across languages in a instance.

\subsection{Notation}\label{sec:model}
Let $D = (I, T_1, T_2, ..., T_K)$ denote a multi-lingual image-text dataset, consisting of the instance $(i_j, t_{j1}, t_{j2}, ..., t_{jK}) \sim D $, where $j$ indexes the instance, $i_j$ is the image in this instance, $t_{jk}$ is the text in the $k$-th language in this instance, and $K$ refers to the total number of languages in the dataset. If it is clear from the context, we will remove the subscript $j$ or $jk$ for brevity. 

\subsection{1-to-K Contrastive Learning}\label{sec:cl}

To solve both two problems in the previous section, the key is that the texts in all languages should be aligned with the semantically similar images all at once. Obviously, it is not possible to do this by aligning pairs of data. Even if uniformly sampling one from the texts in all languages and combining it with the corresponding image to form an image-text pair, the second problem remains. Therefore, the effective way is to form the texts in all languages and the image directly into a tuple as the input. Therefore, we propose a 1-to-K contrastive learning approach to solve this problem. For simplicity, let $\hat{t}$ and $\hat{i}$ represent the normalized text and image representations, respectively. Then, the optimization objective of 1-to-K contrastive learning can be formulated as follows:

\begin{equation}\label{eqn:mcl_i2t}
    \mathcal{L}_{\rm{kcl}}^{\rm{i2t}} = -\frac{1}{K}\log\frac{\exp(\hat{i}_j^T \hat{t}_{jk}/\tau)}{\sum_k^K \exp(\hat{i}_j^T \hat{t}_{jk}/\tau) + \sum_{n,n\neq j}^{N}\sum_k^{K} \exp(\hat{i}_{j}^T \hat{t}_{nk}/\tau)}
\end{equation}

\begin{equation}\label{eqn:mcl_t2i}
    \mathcal{L}_{\rm{kcl}}^{\rm{t2i}} = -\log\frac{\exp(\hat{t}_{jk}^T \hat{i}_j /\tau)}{\exp(\hat{t}_{jk}^T \hat{i}_j /\tau) + \sum_{n, n\neq j}^{N} \exp(\hat{t}_{jk}^T \hat{i}_n /\tau)}
\end{equation}
\noindent
where $K$ is the number of languages and $N$ is the number of negative instances. It is worth noting that there exists literature on multiple positive contrastive learning in other fields \cite{song2020multi,tian2020contrastive}, where all positive items are accumulated in the numerator and the probability of the overall positive terms probability is calculated to be infinitely convergent to 1. Instead, we further set the label of each positive item to 1/K to ensure equal contribution from each language.

Note that increasing the number of multi-lingual texts used as input to the encoders only results in a small increase in GPU memory and training time since the text encoders are usually more lightweight than image encoders in CCP \cite{zeng2022cross} and most of the computations involved are matrix operations that support parallelism. The changes in memory usage and training time before and after applying 1-to-$K$ contrastive learning are detailed in Appendix \ref{sec:appx_gpu}.

\subsection{Pretraining Model: CCR$^k$}\label{sec:pretraining}

Based on the proposed 1-to-K contrastive learning, we further propose a CCP model named CCR$^k$. Specifically, we combine 1-to-K contrastive learning with two other common CCP tasks and balance positive and negative samples by hard sample mining. As shown in the middle of Figure \ref{fig:model}, we adopt the common framework in cross-lingual cross-modal pretraining \cite{ni2021m3p,zhou2021uc2,zeng2022cross}, which consists of a multi-lingual text encoder $f(\cdot)$, a visual encoder $g(\cdot)$ and a fusion encoder $\phi(\cdot,\cdot)$ with image-to-text cross-attention. 

\subsubsection{Hard Sample Mining}
Incorporating cross-attention between the image representation and the text representations in all languages can greatly increase the pre-training time. Therefore, we use the hard sample mining strategy proposed by \citet{li2021align} for both positive and negative samples. This method allows the model can only focus on how to reconstruct the hardest positive samples in the CMLM task and distinguish the hardest negative samples in the MITM task. In subsequent sections, we use $t_j^{\rm{pos}}$ to represent the hard positive sample for texts and $t_j^{\rm{neg}}$ and $i_j^{\rm{neg}}$ to represent the hard negative sample for texts and images, respectively. Please refer to Appendix \ref{sec:appx_hard_neg} for sampling details.

\subsubsection{Multi-lingual Image-Text Matching (MITM)}
The MITM task is a binary classification task that aims to identify whether the semantics of a given image-text pair match. This task is often regarded as an image-text bi-directional prediction problem. Specifically, in the image-to-text direction, the model is trained to select the right one from the hard positive and hard negative text samples. Let $u_{\rm cls}$ be the representation output by the fusion encoder, then the loss function of MITM can be expressed as

\begin{equation}\label{eqn:mitm_i2t}
    \mathcal{L}_{\rm{mitm}}^{\rm{i2t}} = -\log \frac{\exp(\psi(u_{\rm cls}^{\rm p}))}
    {\exp(\psi(u_{\rm cls}^{\rm p}))+\exp(\psi(u_{\rm cls}^{\rm nt}))}
\end{equation}
where $\psi \in \mathbb{R}^{d \times 2}$ is the binary-classification head, $d$ is the representation dimension, $u_{\rm cls}^{\rm p}$ is obtained from $\phi(\hat{t}_j^{\rm pos}, \hat{i}_j)$ and $u_{\rm cls}^{\rm nt}$ is obtained from $\phi(\hat{t}_j^{\rm neg}, \hat{i}_j)$. Similarly, for the text-to-image direction, the matching objective can be expressed as

\begin{equation}
    \mathcal{L}_{\rm{mitm}}^{\rm{t2i}} = -\log \frac{\exp(\psi(u_{\rm cls}^{\rm p}))}{\exp(\psi(u_{\rm cls}^{\rm p}))+\exp(\phi(u_{\rm cls}^{\rm ni})}
\end{equation}
where $\psi \in \mathbb{R}^{d \times 2}$ is the same binary-classification that is used in Eqn. \eqref{eqn:mitm_i2t} and $u_{\rm cls}^{\rm ni}$ is obtained from $\phi(\hat{t}_j^{\rm pos}, \hat{i}_j^{\rm neg})$.

\subsubsection{Cross-Modal Masked Language Modeling (CMLM)}\label{sec:mlm}
The cross-modal masked language modeling task aims to reconstruct the masked tokens using both textual contextual information and image information. Let $t_{j}^{\rm mask}$ be the variant of $t_{j}^{\rm pos}$ whose partial tokens are masked, and $\hat{u}_{j}^{\rm mask}$ is the fusion encoder output corresponding to $t_{j}^{\rm mask}$, then the loss function for this task can be expressed as
\begin{equation}
    \mathcal{L}_{\rm{cmlm}} = -\log \frac{\exp(\rho(\hat{u}_{j}^{\rm mask}, w^+_j))}{\sum_{w_j\in \mathcal{W}}\exp(\rho(\hat{u}_{j}^{\rm mask}, w_j))}
\end{equation}
where $\rho: (\mathbb{R}^d \times \mathcal{W})\rightarrow \mathbb{R}^1$ is a score function to evaluate the matching degree of a given contextual representation with a given token, $w^+_j$ is the original token of the masked location and $\mathcal{W}$ is the vocabulary list. We use the special token [MASK] to replace 15\% of the tokens in each text, following BERT \cite{devlin2018bert}.

\subsubsection{Optimization Objective}\label{sec:opt_objective}
Note that contrastive loss, image-text matching, and masked language modeling have been verified in numerous prior works \cite{li2021align,zeng2022cross} to converge together when co-optimized, so we directly sum them here without the additional hyper-parameters for weighting different losses. Thus, the final optimization objective, which can be expressed as
% Considering the long training time and the fact that all loss values lie on the same order of magnitude, we sum all loss functions mentioned above directly to obtain the final optimization objective, which can be expressed as
\begin{equation}
    \mathcal{L} = \mathcal{L}_{\rm{kcl}}^{\rm{i2t}} + \mathcal{L}_{\rm{kcl}}^{\rm{t2i}} + \mathcal{L}_{\rm{mitm}}^{\rm{i2t}} + \mathcal{L}_{\rm{mitm}}^{\rm{t2i}} + \mathcal{L}_{\rm{cmlm}}
\end{equation}

\subsection{Evaluation Metric: Mean Rank Variation}\label{sec:mrv}
While Recall@K is the common metric used in CCR, it only can reflect the overperformance on a single language. In this section, we introduce a new evaluation metric, Mean Rank Variation (MRV), to measure the rank consistency in different languages within an instance. Figure \ref{fig:model} illustrates the difference between MRV and Recall@K in their calculation methods. MRV for K languages can be computed in both Image-to-Text Retrieval (TR) and Text-to-Image Retrieval (IR) tasks. For example, in the TR task, given an image $i_j$ and a text set in a particular language $\{t_{jk}\}_{j=1}^N$, the similarities between the image and the text set are computed first. Then the text set is sorted by these similarities in ascending order and the rank of $t_{jk}$ is denoted as $Rank_{jk}$.  For each $i_j$, we can loop through $k$ from 1 to $K$ to obtain $\{Rank_{jk}\}_{k=1}^K$, and average them to obtain $\overline{Rank_j}$. Similarly, in the IR task, we denote the rank of retrieving the image $i_j$ using the text $t_{jk}$ as $Rank_{jk}$ and the average rank obtained by retrieving $i_j$ using all $K$ languages as $\overline{Rank_j}$. Thus, MRV for $K$ languages, which is denoted as ${\rm MRV_K}$, can be expressed as
\begin{equation}
    {\rm MRV_K} = \frac{1}{NK}\sum_j^N \sum_k^K {|Rank_{jk} - \overline{Rank_j}|}^2
\end{equation}

Note that there is no trade-off between Recall@K and ${\rm MRV_K}$, which means that when Recall@1=1 holds for all K languages, ${\rm MRV_K}$=0 also holds. ${\rm MRV_K}$ is more likely to reflect the alignment consistency of local semantic space. Such consistency is significant in certain scenarios, such as cross-border e-commerce, to ensure consistency in the results retrieved when the queries are in different languages but have the same semantics.

\section{Experiment}
\subsection{Experiment Setup}
\subsubsection{Pre-training Datasets}\label{sec:pre-traing_data}
For pre-training, we mainly use Conceptual Captions 3M (CC3M) \cite{changpinyo2021conceptual}, which currently has only 1.8 million image-text pairs from the web due to the inaccessibility of image hyperlinks. To verify the scalability of our approach, we further introduce 3 additional cross-modal web datasets, including SBU Caption \cite{ordonez2011im2text}, Visual Genome \cite{krishna2017visual} and COCO \cite{chen2015microsoft}. For the translated version of the texts, we use the 6-language (English, German, French, Czech, Japanese, and Chinese) translated texts in CC3M provided by UC$^2$ \cite{zhou2021uc2} as well as the same 6-language translated texts in the other three datasets, provided by CCLM \cite{zeng2022cross} for fair comparisons. To further verify the generalizability of our method to more languages, we use the M2M-100-large model \cite{fan2021beyond} to translate the English text in the datasets into an additional 4 languages (Spanish, Indonesian, Russian, and Turkish), following \citet{qiu2022}. Therefore, the total number of text languages used for evaluation is 10, which covers all languages in xFlickr\&CO. We plan to open-source these translated texts for research.

\begin{table*}[htp]\small
    \centering
    \caption{Performance comparison on four retrieval datasets, where IR means text-to-Image Retrieval and TR means image-to-Text Retrieval. Consistent with standard evaluation protocols, Recall@1 on xFlickr\&CO, Accuracy on WIT, and average Recall@K with K=1,5,10 on Multi30K and COCO are reported. We only calculate MRV on xFlickr\&CO and Multi30K because there is no one-to-many relationship between images and texts in WIT, whereas the texts in COCO are from different sources.}
    \begin{tabular}{ll|ccc|cc|ccccc|ccc}
        \toprule
        \multirow{2}{*}{\bf{Model}} & \bf{\#Image+} & \multicolumn{3}{c}{\bf{xFlickr\&CO}} & \multicolumn{2}{c}{\bf{WIT}} & \multicolumn{5}{c}{\bf{Multi30K}} & \multicolumn{3}{c}{\bf{COCO}}\\
        & \bf{\#Text} & IR-R@1 & TR-R@1 & MRV$_4$\textcolor{green}{$\downarrow$} & IR & TR & EN & DE & FR & CS & MRV$_4$\textcolor{green}{$\downarrow$} & EN & ZH & JA \\
        \midrule
        \multicolumn{15}{c}{\textit{Fine-tune model on English training set (Zero-Shot)}} \\
        \midrule
        xUNITER \cite{liu2021visually} & 2.7M+100G & 14.04 & 13.51 & 50.60 & 8.72 & 9.81 & - & - & - & - & - & - & - & - \\ %
        M$^3$P \cite{ni2021m3p}  & 3.3M+100G & 12.91 & 11.90 & 54.58 & 8.12 & 9.98 & 87.4 & 58.5 & 46.0 & 36.8 & 15.38 & 88.6 & 53.8 & 56.0\\ %
        UC$^2$ \cite{zhou2021uc2} & 3.3M+19.8M & 20.31 & 17.89 & 21.52 & 7.83 & 9.09 & 87.2 & 74.9 & 74.0 & 67.9 & 6.16 & 88.1 & 82.0 & 71.7\\ %
        TD-MML \cite{qiu2022} & 2.8M+52.0M & 21.30 & 26.35 & - & 9.76 & 10.61 & - & - & - & - & & - & - & - \\ %
        CCLM-3M \cite{zeng2022cross} & 2.8M+54.8M & 64.47 & 62.74 & 13.27 & - & - & 90.4 & 89.9 & 89.4 & 88.1 & 3.18 & 92.3 & 90.4 & 87.3 \\ % cclm reproduce
        \midrule
        CCR$^6$ & 1.8M+10.8M & 29.16 & 28.72 & 15.02 & 6.78 & 8.73 & 88.5 & 87.1 & 87.8 & 85.7 & 3.06 & 91.6 & 89.6 & 86.0 \\
        CCR$^6$-E & 3.3M+19.8M & 32.89 & 33.06 & 7.97 & 6.44 & 8.34 & 90.8 & 90.3 & 91.0 & 89.4 & {\bf 1.28} & 92.5 & 91.4& \textbf{89.4}\\ 
        CCR$^{10}$ & 1.8M+18.0M & 55.45 & 54.88 & 18.96 & 9.94 & 11.73 & 84.2 & 82.5 & 81.9 & 80.8 & 4.13 & 90.0 & 88.0 & 81.7\\
        CCR$^{10}$-E & 3.3M+33.0M & {\bf 73.30} & {\bf 72.64} & {\bf 7.89} & {\bf 11.11} & {\bf 12.62} & {\bf 91.4} & {\bf 90.7} & {\bf 91.3} & {\bf 89.8} & 2.53 & {\bf 92.5} & {\bf 91.4} & 89.3\\
        \midrule
        \multicolumn{7}{c}{\textit{Few-shot fine-tune ``English fine-tuned model'' on target languages (Few-Shot)}} \vline& \multicolumn{8}{c}{\textit{Single-language fine-tune}} \\
        \midrule
        xUNITER & 2.7M+100G & 14.30 & 13.54 & - & - & - & - & - & - & - & - & - & - & - \\
        M$^3$P  & 3.3M+100G & 13.21 & 12.26 & - & - & - & 87.4 & 82.1 & 67.3 & 65.0 & - & 88.6 & 75.8 & 80.1\\
        UC$^2$  & 3.3M+19.8M & 19.79 & 17.59 & - & - & - & 87.2 & 83.8 & 77.6 & 74.2 & - & 88.1 & 84.9 & 87.3\\
        CCLM-3M & 2.8M+54.8M & 65.31 & 63.91 & 12.93 & - & - & 90.4 & 89.6 & 90.0 & 88.8 & 2.41 & 92.3 & 92.1 & 92.4 \\ % cclm reproduce
        \midrule
        CCR$^6$ & 1.8M+10.8M & 29.28 & 28.72 & 15.24 & - & - & 88.5 & 88.1 & 88.6 & 87.3 & 2.28 & 91.6 & 91.0 & 91.8 \\
        CCR$^6$-E & 3.3M+19.8M & 33.19 & 33.41 & 7.81 & - & - & 90.8 & 90.5 & {\bf 91.4} & {\bf 90.5} & 1.30 & 92.5 & {\bf 92.6} & 92.5 \\
        CCR$^{10}$ & 1.8M+18.0M & 55.91 & 55.24 & 18.11 & - & - & 84.2 & 83.6 & 84.6 & 82.5 & 3.90 & 90.0 & 89.7 & 90.4\\
        CCR$^{10}$-E & 3.3M+33.0M & {\bf 73.74} & {\bf 73.27} & {\bf 7.62} & - & - & \bf{91.4} & \bf{90.6} & 91.2 & 90.2 & {\bf 1.12} & \bf{92.5} & 92.5 & \bf{92.5}\\
        \bottomrule
    \end{tabular}
    \label{tab:performance}
\end{table*}

\subsubsection{Baseline}
CCR$^k$ proposed in this paper is mainly an improvement of the training optimization objective in the pre-training phase, so we mainly compare it with other CCP models, including xUNITER \cite{liu2021visually}, UC$^2$ \cite{zhou2021uc2}, M$^3$P \cite{ni2021m3p}, TD-MML \cite{qiu2022} and CCLM \cite{zeng2022cross}. These methods have been briefly described in Section \ref{sec:ccp}, while for more details on them, please refer to Appendix \ref{sec:appx_baseline}.

\subsubsection{The Variant of CCR$^k$}
We report the performance of four model variants pre-trained with different data, which are as follows:
\begin{itemize}
    \item {\bf CCR$^6$} pre-trained using CC3M with 6-language texts.
    \item {\bf CCR$^{10}$} pre-trained using CC3M with 10-language texts.
    \item {\bf CCR$^6$-E} pre-trained using CC3M, COCO, VG and SBU with 6-language texts.
    \item {\bf CCR$^{10}$-E} pre-trained using CC3M, COCO, VG and SBU with 10-language texts.
\end{itemize}

\subsubsection{Evaluation Datasets and Protocols}
We evaluate our methods on four popular CCR datasets, including xFlickr\&CO \cite{bugliarello2022iglue}, WIT \cite{bugliarello2022iglue}, Multi30K \cite{young2014image}, and COCO \cite{chen2015microsoft,li2019coco,yoshikawa2017stair}. Although the images in xFlickr\&CO are derived from the original Flickr30K and COCO, the multi-lingual texts in xFlickr\&CO are manually re-annotated. Therefore, the performance on xFlickr\&CO may not be strongly correlated with that on Multi30K and COCO. For both xFlickr\&CO and WIT, we evaluate our models using two protocols: fine-tuning on the English train set (\textit{Zero-Shot}) and fine-tuning on 100 instances of other languages based on English fine-tuned models (\textit{Few-Shot}). For Multi30K and COCO, we also use two evaluation protocols: fine-tuning on the English train set (\textit{Zero-Shot}) and fine-tuning on each language train set (\textit{Fine-Tune}). Note that the results on WIT under the few-shot scenario are not reported because IGLUE \cite{bugliarello2022iglue} does not provide the corresponding evaluation protocol. For more details, please refer to Appendix \ref{sec:appx_evaluation_dataset}.

\subsection{Implementation Details}

Following \cite{zeng2022cross}, the image encoder is initialized using the 12-layer Swin Transformer \cite{liu2021swin}, and the multi-lingual encoder and fusion encoder are initialized using the pre-trained XLM-R \cite{conneau2020unsupervised}, which consist of 6 layers for each. We provide a detailed comparison of the model architecture and initialization sections between CCR and other baselines in Appendix \ref{sec:appx_baseline}. Also, keeping consistent with \cite{zeng2022cross} for a fair comparison, $\tau$ in Eqn. \eqref{eqn:mcl_i2t} and \eqref{eqn:mcl_t2i} are set as 0.07. The AdamW \cite{loshchilov2019decoupled} optimizer with 1e-4 learning rate, 0.01 weight decay, and first 3\% linearly warm-up steps is used. The batch size on each GPU is set to 64. The pre-training experiments were conducted on 2 NVIDIA A100s, while fine-tuning was done on 1 A100. We pre-train all models for 30 epochs. With the acceleration of PyTorch DDP \cite{li2020pytorch}, it takes approximately 4 days to pre-train for 30 epochs on CC3M with 6 languages. In addition, we provide the hyper-parameters used for fine-tuning all four datasets in Appendix \ref{sec:appx_hyper}.

\subsection{Main Performance}
We report the performance of all four variants of CCR$^k$ and baselines in Table \ref{tab:performance}. Note that the results of CCLM-3M on WIT are not reported in Table \ref{tab:performance} as we find that there is a significant overlap between the WIT test set and the pre-training data of CCLM. Unless otherwise noted, we use ISO 639-1 Abbreviations to represent specific languages in subsequent tables. The table mapping the two-letter codes to the specific language is provided in Appendix \ref{sec:appx_iso639} for convenience.

\paragraph{\bf Recall Rates} With a smaller scale pre-trained data (\#images and \#texts) and fewer language numbers than the baselines, CCR$^{10}$-E achieves SOTA results under both zero-shot and few-shot (or fine-tuning) setting for all CCR datasets, demonstrating the good generalizability and transferability of CCR$^k$ among different languages. When comparing the performance difference among the four variants of CCR$^k$, we can find that (1)  CCR$^{10}$ use more languages compared to CCR$^6$, causing it to improve the performance on the newly added languages while hurting Recall@K of the original languages existing in CCR$^6$, possibly due to the increased difficulty of alignment across more languages; (2)  CCR$^6$-E achieves higher Recall@K and lower MRV on the original languages compared to CCR$^6$ after introducing more pre-training data. 

\paragraph{\bf Consistency Evaluation of Recall@K}
Recall that one of the inconsistency problems leads to inconsistent recall@K in different languages. As seen in Table \ref{tab:performance}, all baselines perform better in English than in other languages on Multi30K and COCO because English is used as a bridge between the visual and other languages during their pre-training. Benefitting from the 1-to-K contrastive paradigm, all four variants of CCR$^k$ maintain significantly smaller inter-language gaps on these two datasets. Among them, CCR$^{10}$-E maintains the smallest performance gap across languages on Multi30K and COCO in the zero-shot scenario, even though this scenario is more favourable for English-related retrieval. More surprisingly, when CCR$^k$ is fine-tuned in each language separately, the performance gap on various languages almost disappears, which reflects the promising application of CCR$^k$ in practical applications. 

\paragraph{\bf Consistency Evaluation of Rank}
Recall that the other problem results in the inconsistency of rank. The motivation behind proposing MRV is that Recall@K cannot reflect such differences across languages within an instance. Therefore, we calculate MRV for four languages (EN, DE, JA, and ZH) on xFlickr\&CO and four languages (EN, DE, FR, and CS) on Multi30K, which are denoted as MRV$_4$ in Table \ref{tab:performance}. We also report MRV$_4$ of all compared models except TD-MML based on the checkpoints obtained from the official IGLUE GitHub repository \footnote{\url{https://github.com/e-bug/iglue}}. It can be found that MRV$_4$ for CCLM, which uses 1-to-1 contrastive learning, has improved substantially compared to M$^3$P and UC$^2$, while CCR$^{k}$ can improve further and achieve the lowest MRV. Similar to Recall@K, adding more languages (CCR$^6$ $\rightarrow$ CCR$^{10}$ and CCR$^6$-E $\rightarrow$ CCR$^{10}$-E) will result in a higher MRV due to the capacity constraints of the model and the elevated difficulty of the optimization objective.

\begin{table}[htp]\small
    \centering
    \caption{Ablation study on pre-training tasks. For Multi30K and COCO, the average of all languages is reported.}
    \begin{tabular}{l|ccccc}
    \toprule
        \multirow{2}{*}{{\bf Model}} & \multicolumn{2}{c}{\bf xFlickr\&CO} & \multicolumn{2}{c}{\bf Multi30K} & {\bf COCO} \\
        & Avg R@1 & MRV$_4$ & Avg Lang & MRV$_4$ & Avg Lang\\
    \midrule
        CCR$^6$ & {\bf 28.94} & {\bf 15.02} & {\bf 87.3} & {\bf 3.06} & {\bf 89.1}  \\
        -w/o KCL & 24.67 & 18.89 & 82.9 & 8.21 & 87.2 \\
        -w/o H-MITM & 27.99 & 15.14 & 85.9 & 4.72 & 88.2 \\
        -w/o H-CMLM & 26.20 & 16.51 & 79.2 & 6.83 & 87.5  \\
    \midrule
        CCR$^{10}$-E & {\bf 72.97} & {\bf 7.89} & {\bf 90.8} & {\bf 2.53} & {\bf 91.1}  \\ 
        -w/o KCL & 68.14 & 11.02 & 84.6 & 6.45 & 88.8 \\ 
        -w/o H-MITM & 70.95 & 8.29 & 87.4 & 3.89 & 90.3 \\
        -w/o H-CMLM & 69.48 & 9.45 & 85.9 & 4.78 & 89.6 \\
    \bottomrule
    \end{tabular}
    \label{tab:ablation}
\end{table}

\subsection{Ablation Study}\label{sec:ablation_study}
To verify the effectiveness of each model component, we conduct ablation experiments by removing critical components. The ablated variants we consider are as follows:
{\bf w/o KCL}: 1-to-K Contrastive Learning (KCL) is replaced with 1-to-1 contrastive learning;
{\bf w/o H-MITM}: Hard sample mining for MITM is replaced with random uniform sampling from the candidate set;
{\bf w/o H-CMLM}: Hard sample mining for CMLM is replaced with uniform sampling from the candidate set.

Due to space constraints, we only report results for CCR$^6$ and CCR$^{10}$-E under the zero-shot setting in Table \ref{tab:ablation}. Note that the other two variants also show a similar trend. As can be seen from the results, each pre-training task and sampling approach proposed to contribute to the improvement in both Recall@K and MRV$_4$. More specifically, 1-to-K contrastive learning has the largest improvement for all metrics, while 1-to-1 contrastive learning is still better than the results without contrastive learning. Hard sample mining positively affected both MITM and CMLM downstream tasks.

\subsection{Further Study}

\subsubsection{Pure Contrastive Learning}\label{sec:pure_contrastive_learning}
\begin{figure}[htb]
\centering  %图片全局居中
\centering  %图片全局居中
\subfigure[The comparison of pure contrastive learning on XFlickr\&Co.]{
\includegraphics[width=0.48\linewidth]{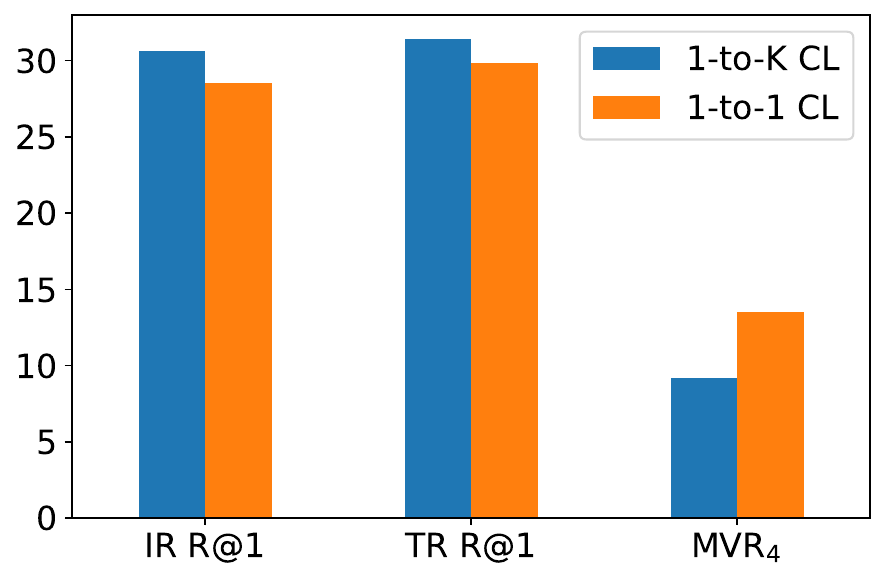}
\label{fig:12k_vs_121}}
\subfigure[The comparison of pure contrastive learning on Multi30K.]{
\includegraphics[width=0.48\linewidth]{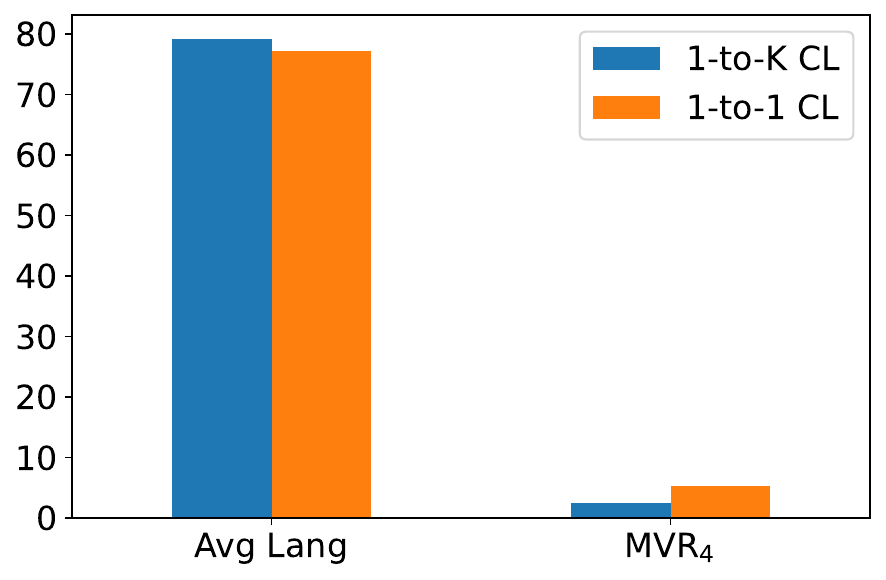}} \\
\subfigure[Comparison of loss function value and average Recall@K with K=1,5,10 on Multi30K when using 1-to-1 contrastive learning and 1-to-K contrastive learning.]{
\includegraphics[width=\linewidth]{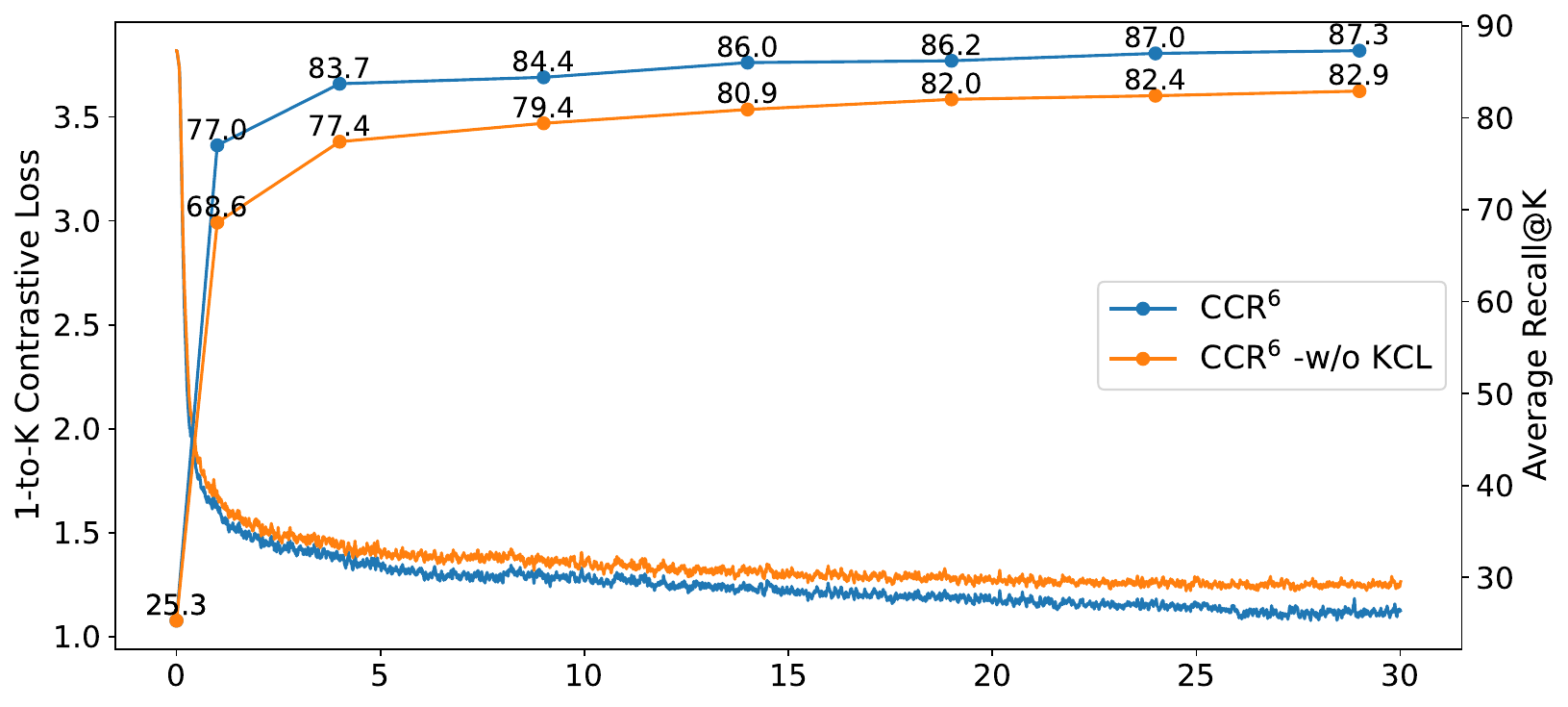}
\label{fig:loss_and_recall}} \\
\subfigure[T-SNE visualization of CCR$^6$.]{
\includegraphics[width=0.47\linewidth]{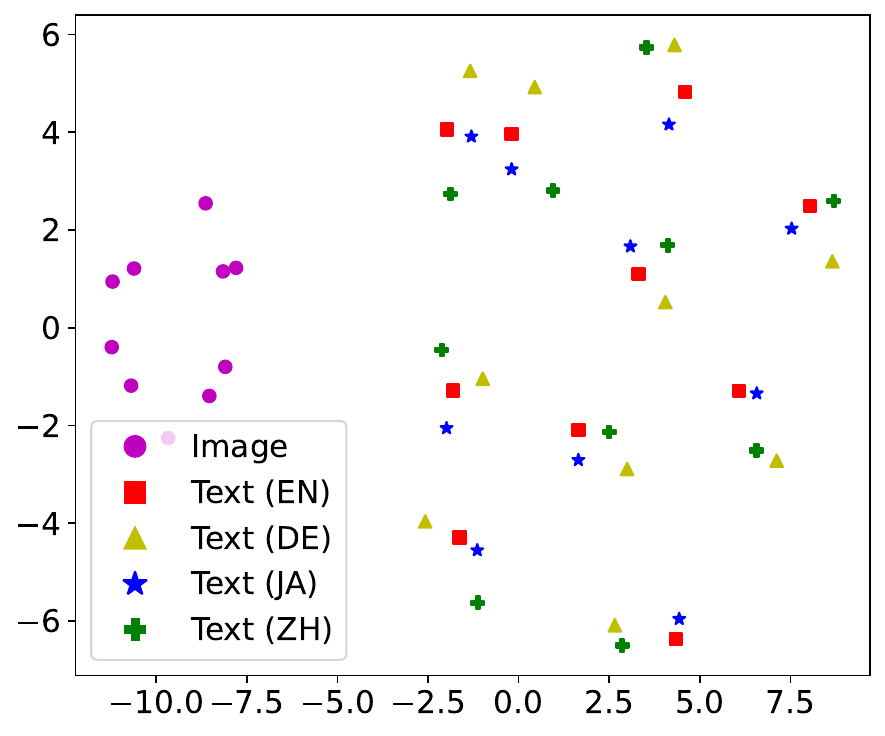}
\label{fig:t_sne_12k}}
\subfigure[T-SNE visualization of CCR$^6$ -w/o KCL.]{
\includegraphics[width=0.47\linewidth]{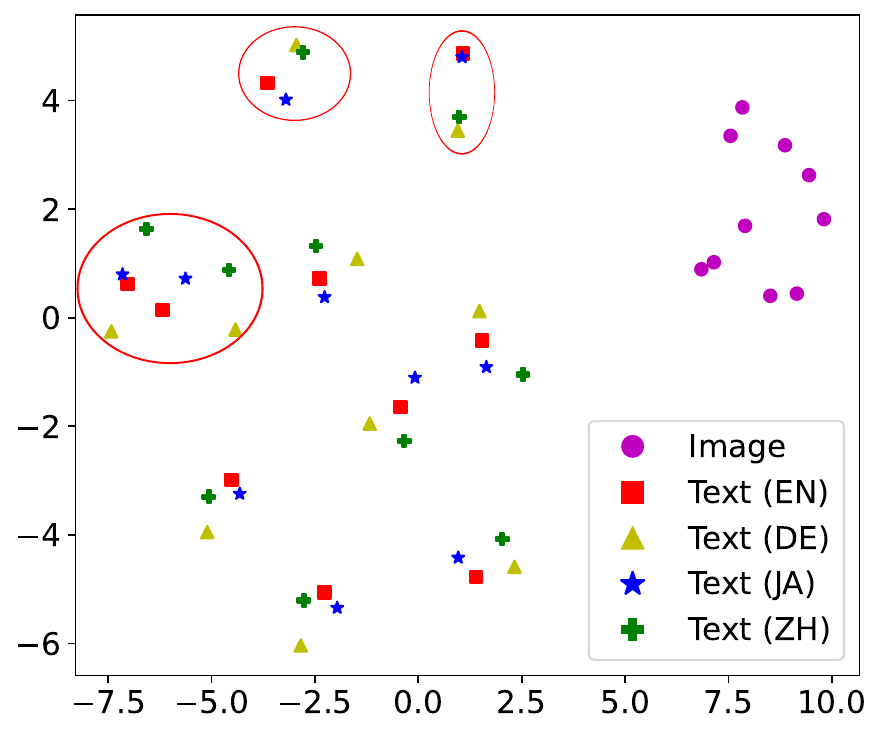}
\label{fig:t_sne_121}}
\caption{Futher Study in Alignment Process.}
\end{figure}

In fact, CCR$^k$ is proposed to ensure that the model's parameter number and pre-training tasks are similar to other baselines. However, neither MITM and CMLM tasks nor the fusion encoder is necessary for the retrieval task. Therefore, we further compare the effect of 1-to-K and 1-to-1 contrastive learning on Recall@K and MRV with the fusion encoder removed, while other settings remain consistent with CCR$^6$. As seen from Figure \ref{fig:12k_vs_121}, 1-to-K contrastive learning can still lead on both xFlickr\&CO and Multi30k.

\subsubsection{Loss and Performance}
To better understand why our method works, we record the 1-to-1 contrastive loss and 1-to-K contrastive loss during the pre-training process of ``CCR$^6$'' and ``CCR$^6$ -w/o KCL'', respectively. In addition, we evaluate the checkpoints every 5 epochs on Multi30K under zero-shot setting and plot the results in Figure \ref{fig:loss_and_recall}. The figure shows that 1-to-K contrastive learning performs better at all evaluated checkpoints. Attributed to the absence of directional bias, when pre-training with 1-to-K contrastive learning, the corresponding loss values remain lower than those when using 1-to-1 contrastive learning.

\subsubsection{T-SNE Visualization} A T-SNE visualization similar to that in Section \ref{sec:direction_bias} is shown in Figure \ref{fig:t_sne_12k} and Figure \ref{fig:t_sne_121}, which contains 10 instances randomly sampled in xFlickr\&CO. Comparing to 1-to-1 contrastive learning, 1-to-K contrastive learning enables higher discrimination between instances and a more balanced distribution within instances. In addition, a case study on failure alignment is provided in Appendix \ref{sec:case_study} for potential further improvement.

\begin{figure*}[ht]
    \centering
    \includegraphics[width=\linewidth]{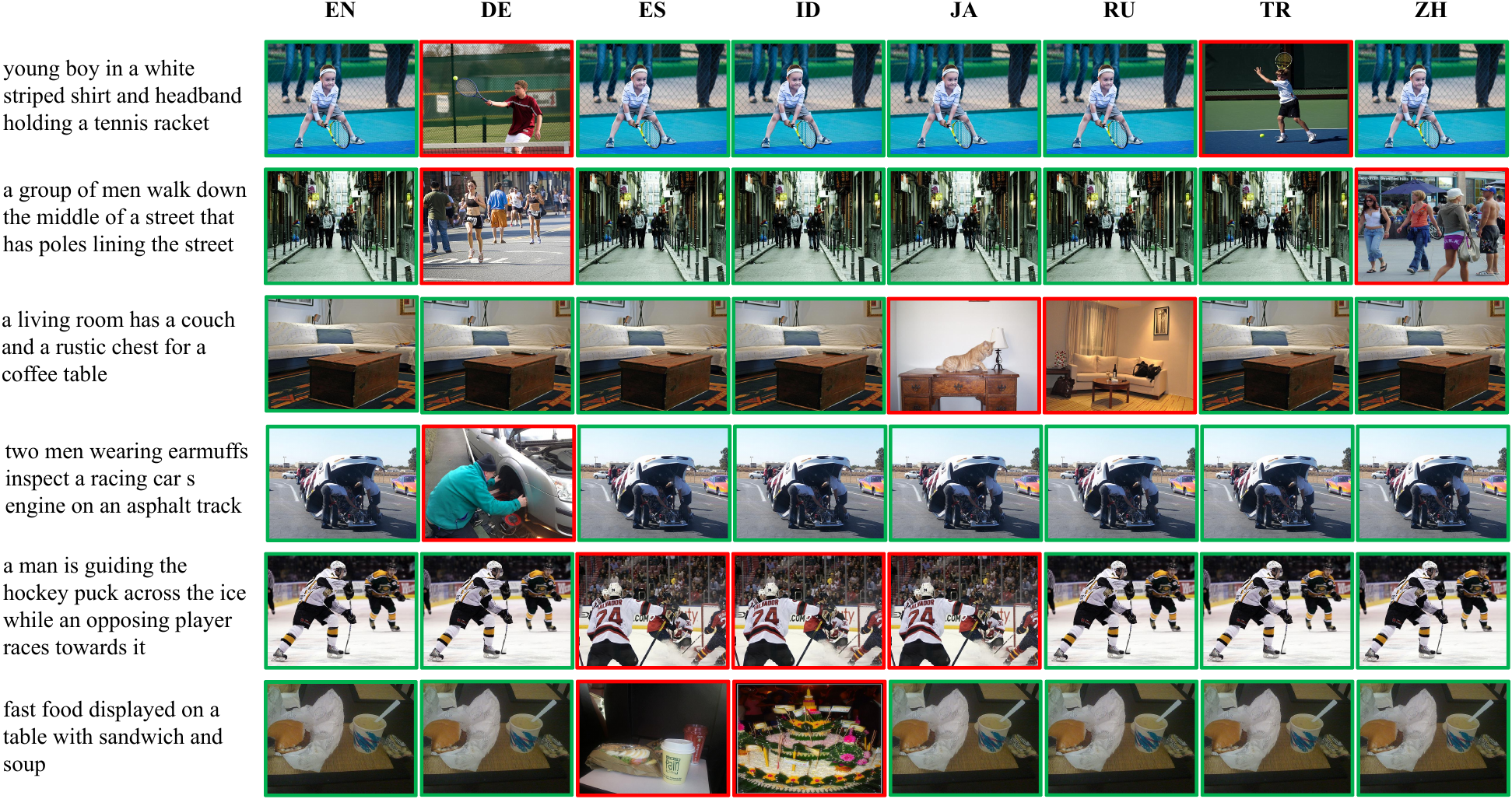}
    \caption{Six wrong cases of text-to-Image Retrieval (IR) on xFlickr\&CO. We only provide the English text in each instance as a reference, and the images are actually retrieved from the text corresponding to the labelled language at the top of each column. The green and red boxes outside the images represent the correct and incorrect images.}
    \label{fig:case_study}
\end{figure*}

\section{Case Study}\label{sec:case_study}
After manually analyzing the wrong cases in xFlickr\&CO, which are not correct under some language settings, we summarized two typical causes of matching errors: fine-grained semantic matching errors and pseudo-negative samples. We give some cases for each of them in Figure \ref{fig:case_study}. Since images are more presentable and comprehensible than texts, we only use the error cases from the text-to-image retrieval (IR) task. The first four cases demonstrate a fine-grained semantic matching error. For example, the concept of ``headband'' in the first case is so specialized that the image can match all other features when retrieved using German (DE) and Turkish (TR). The last two cases show a pseudo-negative sample error, where the images retrieved actually match the text semantics, but these matching relationships are missing annotations in the dataset. For example, in the fifth case, both images retrieved for the "hockey game" matched the textual description, yet only one is labelled as correct in the xFlickr\&CO dataset.

\section{Discussion}
\paragraph{The Novelty of 1-to-K Contrastive Learning} 
The proposed modification is not groundbreaking but based on traditional 1-to-1 contrastive learning. However, recall that 1-to-1 contrastive learning, which has been carried over from the cross-lingual or cross-modal domains, is still the dominant paradigm in CCP. The call to change a task's pre-training paradigm is usually tough. Changing to 1-to-K contrastive learning is minimal yet effective and easily applicable to the existing CCR models based on SimSiam networks.

\paragraph{The Significance of the Consistency in CCR}
Maintaining consistency in CCR is important. For example, in a cross-border e-commerce business, consistency in recall across languages ensures that the entire retrieval system can be supported by a single fundamental model. Further, the query with the same semantics issued by different native-speaking customers should be expected to return the same results, meaning there needs to be good consistency in rank across different languages within an instance. If we evaluate the retrieval model with Recall@K on each language only, the true performance of the CCR model will not be reflected.

\paragraph{Further Consistency}
Ensuring equal contributions across languages in all aspects is challenging. For instance, XLM-R, CCR$^k$'s cross-lingual encoder, is trained on the 2.5TB CommonCrawl Corpus encompassing 100 languages. Discrepancies in data sizes between high-resource and low-resource languages within this corpus, like the 100GB English data versus the 0.1GB Sundanese data, impede XLM-R from achieving uniform performance across languages. Balancing language contributions during pre-training could help narrow the performance gap but would require substantial computational resources, which we will explore in future studies.

\section{Conclusion}
In this paper, we first analyze the two problems of inconsistency existing in the current CCP methods and point out their impact on CCR via theoretical analysis and empirical studies. Then we propose a 1-to-K contrastive paradigm and a CCP model, CCR$^k$, based on it, which equally aligns all languages with vision at once, effectively improving the consistency in CCR. In addition, a new evaluation metric, MRV, is proposed to portray the consistency of each language rank within each instance. Exclusive experiments on the four CCR datasets show that our model scales well and achieves new SOTA on both Recall@K and MRV.

\section*{Acknowledgements}
This work was supported by the National Science and Technology Major Project under Grant 2022ZD0120202, in part by the National Natural Science Foundation of China (No. U23B2056), in part by the Fundamental Research Funds for the Central Universities, and in part by the State Key Laboratory of Complex \& Critical Software Environment.

\bibliographystyle{ACM-Reference-Format}
\bibliography{sample-base}

\appendix
\section{ISO 639 Language Codes}\label{sec:appx_iso639}
We give the ISO-691 codes for all the language codes that appear in the main text and appendices in Table \ref{tab:iso639} for reference.
% language ISO 639-1
\begin{table}[htp]
\centering
\caption{Part of codes and languages in ISO 639-1.}
\begin{tabular}{c|ccc}
\toprule
\bf{Code} & \bf{Language} & \bf{Family} & \bf{Script} \\
\midrule
AR & Arabic & Afro-A & Arabic \\
BG & Bulgarian & Indo-E & Cyrillic \\
CS & Czech & Indo-E & Latin \\
DA & Danish & Indo-E & Latin \\
DE & German & Indo-E & Latin \\
EL & Greek & Indo-E & Greek\\
EN & English & Indo-E & Latin \\
ES & Spanish & Indo-E & Latin \\
ET & Estonian & Uralic & Latin \\
FR & French & Indo-E & Latin \\
ID & Indonesian & Austron & Latin \\
JA & Japanese & Japonic & Kanji \\
KO & Korean & Koreanic & Hangul \\
RU & Russian & Indo-E & Cyrillic \\
TR & Turkish & Turkic & Latin \\
VI & Vietnamese & Austro-A & Latin \\
ZH & Chinese & Sino-T & Hanzi \\
\bottomrule
\end{tabular}
\label{tab:iso639}
\end{table}

\section{Supplement on Experiment Setup}\label{sec:appx_setup}
\subsection{Baseline}\label{sec:appx_baseline}
This section details the baselines used for comparison and compares key information about their architectures and pre-training processes in Table \ref{tab:baseline_comparsion}.

\paragraph{\underline{\bf xUNITER \cite{liu2021visually}}} is a multi-lingual variant of UNITER \cite{chen2020uniter}, which follows the architecture of UNITER and the parameters are initialized with XLM-R$_{\rm base}$ \cite{conneau2020unsupervised}. It also has a twin, mUNITER, which is initialized using mBERT \cite{devlin2018bert}. Considering that xUNITER works better, we ignore the results of mUNITER in this paper. xUNITER and mUNITER are pre-trained using image-English text pairs and parallel corpus alternately composed of batch.
\paragraph{\underline{\bf UC$^2$ \cite{zhou2021uc2}}}  presents the first MT-augmented pre-training model that pivots primarily on images and complementary on English to learn cross-lingual cross-modal representation from large-scale of multi-lingual image-to-text pairs. Two new pre-training tasks, Masked Region-to-Token Language Modeling and Visual Translation Language Modeling, are proposed to facilitate the model to obtain better alignment between vision and different languages.
\paragraph{\underline{\bf M$^3$P \cite{ni2021m3p}}} combines multi-lingual pre-training and multi-modal pre-training into a unified framework via multitask Learning. multi-modal code-switched training is proposed to further alleviate the issue of lacking enough labeled data for non-English multi-modal tasks and avoid the tendency to model the relationship between vision and English text.
\paragraph{\underline{\bf TD-MML \cite{qiu2022}}}  uses translated data for multi-lingual multi-modal learning, which are applied in both pre-training and fine-tuning data with the existing CCP model. In order to prevent the model from learning from low-quality translated texts, two metrics are proposed for automatically removing the low-quality translation texts from the resulting datasets.
\paragraph{\underline{\bf CCLM \cite{zeng2022cross}}}  is a CCP framework that unifies cross-lingual pretraining and cross-modal pretraining with shared architectures and objectives. Contrastive learning is introduced for cross-modal and cross-lingual alignment, respectively.

\begin{table}[htp]\footnotesize
    \centering
    \caption{The image feature source, backbone initialization method, and the language number (\#Lang) involved in pre-training for each CCP model.}
    \begin{tabular}{l|ccc}
    \toprule
       {\bf Model} & {\bf Image Feature Source} & {\bf Initialization of Backbone} & {\bf \#Lang} \\
    \midrule
       \multirow{2}{*}{xUNITER} & 36 Rols from Faster & XLM-R$_{\rm base}$ & \multirow{2}{*}{104} \\
       & R-CNN with ResNet-101 & (12 layers) \\
    \midrule
       \multirow{2}{*}{UC$^2$} & 36 Rols from Faster & XLM-R$_{\rm base}$ & \multirow{2}{*}{104} \\
       & R-CNN with ResNet-101 & (12 layers) \\
    \midrule
       \multirow{2}{*}{M$^3$P} & 10-100 Rols from Faster & XLM-R$_{\rm base}$ & \multirow{2}{*}{6} \\
       & R-CNN with ResNet-101 & (12 layers) \\
    \midrule
       \multirow{2}{*}{TD-MML} & 36 Rols from Faster & XLM-R$_{\rm base}$ & \multirow{2}{*}{20} \\
       & R-CNN with ResNet-101 & (12 layers) &  \\
    \midrule
        \multirow{2}{*}{CCLM} & Swin Transformer & Odd-numbered layers & \multirow{2}{*}{20} \\
        & (12 layers, Trainable) & in XLM-R$_{\rm large}$ (12 layers) \\
    \midrule
       \multirow{2}{*}{CCR$^k$} & Swin Transformer & Odd-numbered layers & \multirow{2}{*}{6-10} \\
        & (12 layers, Trainable) & in XLM-R$_{\rm large}$ (12 layers) \\
    \bottomrule
    \end{tabular}
    \label{tab:baseline_comparsion}
\end{table}

\subsection{Evaluation Dataset}\label{sec:appx_evaluation_dataset}
\paragraph{\underline{\bf xFlickr\&CO}} is a novel dataset purposed by ICLUE \cite{bugliarello2022iglue} and collected by combining 1000 images from Flickr30K and COCO respectively. The existing captions from \cite{chen2015microsoft} and \cite{karpathy2015deep} are used for English and Japanese, while the captions are from crowd-source for the other 6 languages.
\paragraph{\underline{\bf WIT}} means ``Wikipedia-based Image-Text'' dataset \cite{srinivasan2021wit} collected instances from the websites of Wikipedia in 108 languages. For training, a subset of 500K captions is randomly sampled from the English training set of WIT. For evaluation, the WIT test data released as part of its corresponding Kaggle competition\footnote{\url{www.kaggle.com/c/wikipedia-image-caption}} is used.
\paragraph{\underline{\bf Multi30K}} extends Flickr30K \cite{young2014image} from English to German, French and Czech. It contains 31,783 images obtained from Flickr and provides five captions per image in English and German, and one caption per image in French and Czech. Dataset splits are defined as the original Flickr30K.
\paragraph{\underline{\bf COCO}} extends the original COCO Caption \cite{chen2015microsoft} by translating the captions into Japanese and Chinese. The Japanese and Chinese subsets consist of 820k and 20k captions respectively. Following previous work, we use the same train, dev, and test splits for English and Japanese as defined by \citet{karpathy2015deep}. For Chinese, we use the COCO-CN split \cite{li2019coco}.

\begin{table}[htp]\small
    \centering
    \caption{Statistics on the datasets for evaluation.}
    \begin{tabular}{c|ccccc}
    \toprule
        \multirow{2}{*}{\bf Dataset} & \multicolumn{2}{c}{\bf Train} & \multicolumn{2}{c}{\bf Test} & \multirow{2}{*}{\bf Language} \\
        & \#Text & \#Image & \#Text & \#Image & \\
    \midrule
        \multirow{2}{*}{xFlickr\&CO} &  \multirow{2}{*}{145K} & \multirow{2}{*}{29K} & \multirow{2}{*}{2K} & \multirow{2}{*}{2K} & DE EN ES ID\\
        & & & & & JA RU TR ZH\\
    \midrule
        \multirow{3}{*}{WIT} & \multirow{3}{*}{500K} & \multirow{3}{*}{469K} & \multirow{3}{*}{9.6K} & \multirow{3}{*}{6.2K} & AR BG DA EL\\
        & & & & & EN ET ID JA \\
        & & & & & KO TR VI \\
    \midrule
        Multi30K & 29K & 29K & 1K & 1K & EN DE FR CS\\
    \midrule
        \multirow{2}{*}{COCO} & 567K & 113K & 25K & 5K & EN JA\\
        & 18K & 18K & 1K & 1K & ZH \\
    \bottomrule
    \end{tabular}
    \label{tab:dataset}
\end{table}

\section{Implementation Details}

\begin{table}[tp]
    \centering
    \caption{Hyper-parameters under the zero-shot settings.}
    \begin{tabular}{l|cccc}
    \toprule
        \bf Parameter & \bf xFlickr\&CO & \bf WIT & \bf Multi30K & \bf COCO \\
    \midrule
        Learning rate & 1e-5 & 3e-5 & 3e-5 &  3e-5 \\ 
        Batch size & 64 & 80 & 64 & 64 \\ 
        Epochs & 10 & 10 & 10 & 10 \\ 
        Max input length & 80 & 80 & 40 & 40  \\ 
    \bottomrule
    \end{tabular}
    \label{tab:hyperpara_zero}
\end{table}

\begin{table}[tp]
    \centering
    \caption{Hyper-parameters under the fine-tuning settings.}
    \begin{tabular}{l|ccc}
    \toprule
        \bf Parameter & \bf xFlickr\&CO & \bf Multi30K & \bf COCO\\
    \midrule
        Shot number & 100 & - & -\\
        Learning rate & 1e-5 &3e-5 &  3e-5\\
        Batch size & 64 & 64 & 64\\
        Epochs & 60 & 10 & 10\\ 
        Max input length & 80 & 40 & 40 \\ 
    \bottomrule
    \end{tabular}
    \label{tab:hyperpara_few}
\end{table}

\subsection{Evaluation Protocols}
\paragraph{\underline{\bf Zero-Shot}} Only pre-training and fine-tuning on the English train set, then evaluate the test set in each target language.
\paragraph{\underline{\bf Few-Shot Fine-tune}} First pre-training and fine-tuning on English train set. Then twice fine-tuning 100 labeled instances in a target language and evaluating the test set of this target language.
\paragraph{\underline{\bf Single-Language Fine-tune}} First pre-training and fine-tuning on English train set. Then, fine-tuning the training set of the target language and evaluating the test set of this target language.

\subsection{Hyperparameter Setting}\label{sec:appx_hyper}

For zero-shot xFlickr\&CO and WIT, we first fine-tune the model on the English training set, and then evaluate zero-shot and few-shot performance in other languages. Following \cite{zeng2022cross}, for both zero-shot and few-shot experiments, we use AdamW optimizer with $\beta_1$ = 0.9 and $\beta_2$ = 0.999; weight decay is set to 0.01; learning rate scheduler is linear. The all hyper-parameters used are shown in Table \ref{tab:hyperpara_zero}.

\subsection{The Method of Hard Negative Sampling}\label{sec:appx_hard_neg}

For positive samples, given an image $i_j$,  its associated set of texts ($t_{j1}, t_{j2}, ..., t_{jK}$) can be regarded as positive samples. Among these texts, the hardest positive sample $t_{ik^{\rm{pos}}}$ can be identified as the text that aligns worst with the image, and the degree of alignment can be estimated by computing the cosine similarity between the image and text representations. Accordingly, we can sample the index $k^{\rm{pos}}$ of the hardest positive sample from a specific distribution $T$, which can be expressed as
\begin{equation}
t_j^{\rm{pos}} = t_{ik^{\rm{pos}}}, \ k^{\rm{pos}} \sim T, where\ P_T(k) = 1-\frac{\hat{t}_{jk}^T \hat{i}_j}{\sum_{k'}^K\hat{t}_{jk'}^T \hat{i}_j}
\end{equation}
where $T$ is a multinomial distribution. 

For negative samples, if the image and the text from different tuples are well aligned, they can be regarded as hard negative samples for each other. Also, we estimate the degree of alignment using the cosine similarity and sample the index of the negative example from a multinomial distribution. Thus, the process of obtaining the hard negative image can be expressed as

\begin{equation}
    i_j^{\rm{neg}} = i_{j^{\rm{neg}}}, \ j^{\rm{neg}} \sim R, 
    where\ P_R(j') =
       \frac{\sum_{k}^K\hat{t}_{jk}^T \hat{i}_j'}{\sum_{j' \neq j}^{N}\sum_{k}^K\hat{t}_{jk}^T \hat{i}_{j'}} \\
\end{equation}
where $R$ is a multinomial distribution. Similarly, we can obtain the hard negative text for each image in the batch.

\subsection{The Method of Rank}
We obtain the representations from the text encoder and image encoder outputs and rank the candidates by cosine similarity. For CCR$^k$ and ablation models containing the fusion encoder, we re-rank only the top $N$ candidates using the Fusion encoder to better adapt to the web-scale data. Specifically, we use the projection head used for the multi-lingual image-text matching task to predict the match probability between the query and each shortlisted candidate and re-rank the candidates regarding this probability only. In our experiment, $N$ is 256 for COCO and 128 for the other three datasets.

\begin{table}[tp]
    \centering
    \caption{Time and memory comparison.}
    \begin{tabular}{l|cc}
    \toprule
        Model &  Training Time (Per Epoch) & Memory (Per A100)\\
    \midrule
        CCR$^1$ & 137min & 27,814MB \\
        CCR$^6$ & 158min (15\%$\uparrow$) & 31,364MB (13\%$\uparrow$) \\
        CCR$^{10}$ & 173min (26\%$\uparrow$) & 34,203MB (23\%$\uparrow$)\\
    \bottomrule
    \end{tabular}
    \label{tab:mermory}
\end{table}

\section{Time and Memory Comparison}\label{sec:appx_gpu}
We compare the model's training time and GPU memory consumption for different language numbers of translated texts, which are reported in Table \ref{tab:mermory}. The results in the table are the average results measured while keeping other external conditions constant as much as possible. It is easy to find that both training time and memory usage increase linearly with the number of languages. Specifically, the training time increases by 4.2 min per language for 1 Epoch, while the memory footprint increases by 710 MB per language per Nvidia A100 40GB.

\end{document}